\documentstyle[12pt]{article}
\begin{document}
\begin{center}
{\Large Differential geometric and topological methods with MHD and plasma physics constraints}
\vspace{1cm}

\noindent
L.C. Garcia de Andrade\footnote{Departamento de Fisica Teorica-Instituto de F\'{\i}sica , Universidade do Estado do Rio de Janeiro-UERJ, Rua S\~{a}o
Francisco Xavier 524, Rio de Janeiro Brasil.E-mail:garcia@dft.if.uerj.br}
\end{center}
\vspace{2cm}
\begin{center}
{\Large Abstract}
\end{center}
\vspace{0.5cm}
Non-solitonic examples of the application of geometrical and topological methods in plasma physics and magnetohydrodynamics (MHD) are given. The first example considers the generalization of magnetic helicity to gravitational torsion loop. The second example considers also the application of this same torsion loop metric to some problems of helical fields in MHD dynamo theory. In the last example a Riemannian magnetic metric is given where the magnetic field itself is present in the diagonal time-independent metric. In this example the MHD equations are shown to be compatible to the geometrical Bianchi identity by making use of Cartan's differential calculus formalism. The Riemann curvature of the helical flow is also obtained.
\newpage
\section{Introduction}
In recent years several examples \cite{1,2,3,4} of the use of differential geometrical and topological methods have been presented in  the literature, ranging from the Schief's generation of toroidal flux surfaces in MHD via soliton theory \cite{1} and the hidden integrability in ideal MHD using the Pohlmeyer-Lund-Regge to the solar physics magnetic topology applications considered by Moffatt and Ricca \cite{3} and Field and Berger \cite{4}. In all these example the differential geometry of curves and surfaces \cite{5} were used. Earlier Sivaram and Garcia de Andrade \cite{6} have used non-Riemannian geometry with torsion to investigate the Debye problem in plasma physics. Those previous work using soliton theory in MHD of course made use of the Riemannian geometry of surfaces where the only torsion that appears was the Serret-Frenet scalar torsion and not the higher-dimensional \cite{7} Élie Cartan \cite{8} torsion with the exception of reference \cite{6}. In this paper we strike back to the non-Riemannian geometry endowed with Cartan vectorial torsion to investigate torsion loops , previously investigated by Letelier \cite{8} in the context of gravitational physics such as Einstein-Cartan gravity, now also in the context of MHD, by substituting the vector field in this metric by the magnetic field itself. We also show that it is possible to generalised the magnetic-helicity topological aspects of plasma physics can be carried over to gravitational physics. Another interesting example is the diagonal Riemannian magnetic metric. This spacetime metric instead of satisfying the Einstein equations ,or either Einstein-Cartan equations in the case of non-Riemannian loops, it fulfills the MHD equations, that is the reason why be called magnetic metric. The magnetic metric also constrain the Bianchi identity via MHD dynamo equation \cite{9,10}. The ideas applied here in the context of MHD and plasma physics keep resemblance with the analog gravity models which compare fluid mechanics and condensed matter systems such as BEC to the Riemannian \cite{11} and non-Riemannian \cite{12} acoustic metrics. The basic difference is that here we use a magnetic effective metric instead an acoustic one. Effective metrics in MHD maybe constructed, in near future by making use of the Navier-Stokes equation of viscous flows with nonlinear magnetic fields term. More recently we have shown that the non-Riemannian structure called acoustic torsion may also exist in this same case. The paper is organised as follows : In the section 2 we present the magnetic topology generalisation to gravitational torsion  loops. In this section we also deal with the non-Riemannian loop magnetic metric constrained by the MHD equations of dynamo theory. Section 3 shows that the more simple Riemannian geometry can take care of the MHD equations by making use of a time-independent diagonal metric. In section 4 discussions and conclusions are presented. 
\section{Magnetic-like topology of torsion loops}
The Letelier teleparallel torsion loop metric is given by
equation 
\begin{equation}
ds^{2}= (dt+\vec{B}.d\vec{x})^{2}- {d\vec{x}}^{2}
\label{1}
\end{equation}
when this metric is used on the Cartan's structure equation of the differential forms a vector relationship similar to the magnetic field derivative appears as
\begin{equation}
\vec{J}= {\nabla}{\times}\vec{B}
\label{2}
\end{equation}
where $\vec{J}$ is the gravitational analog to the magnetic field while $\vec{B}$ is the gravitational analog of the vector potential. It is clear that in analogy of magnetism we here also possess a vector gauge field where $\vec{B}\rightarrow \vec{B}+{\nabla}{\epsilon}$. From expression $(\ref{2})$ we note that the torsion curve is really closed since the torsion vector field is divergence-free : ${\nabla}.\vec{J}=0$. Of course now one can easily show that by defining the gravitational helicity by 
\begin{equation}
H_{g}= \int{\vec{B}.\vec{J}d^{3}x}
\label{3}
\end{equation}
By analogy with the Berger and Field work \cite{1} one has for the change in H according to the gauge freedom transformation that
\begin{equation}
{\Delta}H_{g}= \int{{\nabla}{\epsilon}.\vec{J}d^{3}x}=\int{{\nabla}.({\epsilon}\vec{J})d^{3}x}=\int{{\epsilon}\vec{J}.\vec{n}dS}
\label{4}
\end{equation}
Where the integration of the first two integrals is on total volume of space V otherwise the field lines would close outside V. The Stokes theorem has also been used to obtain the last integral. Note, however, that only when the torsion vector is orthogonal to the vector $\vec{n}$ the helicity is conserved. To try to remedy this situation in the next section we drop the teleparallelism condition ${R^{\mu}}_{{\alpha}{\beta}{\gamma}}=0$ (here ${\alpha},{\beta}$ represents the four-dimensional spacetime coordinates) and consider the computation of other torsion components where the helicity of the metric appears explicitly. Although the spacetime metric $(\ref{1})$ is invariant with respect to a general coordinate transformation as should be the line elements in Riemannian geometry, it is not invariant with respect to above gauge transformations, and the torsion loop metric becomes
\begin{equation}
g_{00}=1
\label{5}
\end{equation}
\begin{equation}
g_{0i}= B_{i}+{\partial}_{i}{\epsilon} 
\label{6}
\end{equation}
\begin{equation}
g_{ij}=-{\delta}_{ij}+B_{i}B_{j}
\label{7}
\end{equation}
where here ${{\nu},{\mu}=0,1,2,3}$ and latin indices takes values from one to three. It is clear that this metric exhibit explicitely the gauge freedom scalar. From the metric components is easy to compute the following components of the Cartan torsion tensor 
\begin{equation}
T_{i0j}=\frac{1}{2}[{\partial}_{i}g_{0j}-{\partial}_{j}g_{0i}]=\frac{1}{2}[{\partial}_{i}B_{j}-{\partial}_{j}B_{i}]
\label{8}
\end{equation}
This expression can be recast in a more ellegant form by writing it in vector form as
\begin{equation}
{\epsilon}^{lij}T_{i0j}=\frac{1}{2}[{\nabla}{\times}\vec{B}]^{l}
\label{9}
\end{equation}
and 
\begin{equation}
T_{i0i}= 0
\label{10}
\end{equation}
where the Einstein summation convention was used in this last expression. Finally the last component of Cartan torsion is
\begin{equation}
{\epsilon}^{lkj}T_{kij}=\frac{1}{2}[{\partial}_{i}{\epsilon}({\nabla}{\times}{\vec{B}})^{l}]
\label{11}
\end{equation}
where ${\epsilon}^{kli}$ is the Levi-Civita symbol. From this last expresion we not that by contracting the indices $l=i$ we note that a new generalised definition of gravitational helicity can be obtained since
\begin{equation}
\int{{\epsilon}^{ikj}T_{kij}d^{3}x}=\frac{1}{2}\int{{\nabla}{\epsilon}.{\nabla}{\times}{\vec{B}}d^{3}x}= H_{g}
\label{12}
\end{equation}
which shows that this new definition coincides with the old with the advantage that now the full torsion tensor is consider and not only the torsion vector part. Since the component $T_{i0j}$ can also be expressed in terms of the vector $\vec{J}$ one may express the gravitational helicity by yet another integral as
\begin{equation}
\int{{\epsilon}^{lij}B_{l}T_{i0j}d^{3}x}=\frac{1}{2}\int{[\vec{B}.{\nabla}{\times}\vec{B}]d^{3}x}= H_{g}
\label{13}
\end{equation}
In the next section we propose a solution for the problem of helicity in Riemann-Cartan spacetime. In this section we show that it is possible to show that gravitational helicity is conserved as long as we extend the spacetime to a more general Riemann-Cartan one instead of the teleparallel spacetime. It is easy to show that  by considering the torsion loop metric (\ref{1}) in differential forms notation 
\begin{equation}
ds^{2}=({\omega}^{0})^{2}-({\omega}^{1})^{2}-({\omega}^{2})^{2}-({\omega}^{3})^{2}
\label{14}
\end{equation}
where the basis one-forms are
\begin{equation}
{\omega}^{0}=(dt+\vec{B}.d\vec{x})
\label{15}
\end{equation}
and ${\omega}^{1}=dx$, ${\omega}^{2}=dy$ and ${\omega}^{3}=dz$. Now a small perturbation of connection one-forms in the teleparallel case according to the formula 
\begin{equation}
{{\omega}^{i}}_{k}= -\frac{1}{2}{{\epsilon}^{in}}_{lp}[{\vec{J}}_{0}]_{m}{\omega}^{n}
\label{16}
\end{equation}
where the new torsion vector field now in RC spacetime ${\vec{J}}_{0}$ is given in terms of the old teleparallel vector field by
\begin{equation}
{\vec{J}}_{0}= \vec{J}- {\nabla}{\times}\vec{B}
\label{17}
\end{equation}
Note from this expression that the vector torsion field ${\vec{J}}_{0}$ represents also a loop since
\begin{equation}
{\nabla}.{\vec{J}}_{0}= {\nabla}.\vec{J}=0
\label{18}
\end{equation}
where we have used the fact that ${\nabla}.[{\nabla}{\times}\vec{B}]=0$. Note also that now the new definition of helicity similar to the previous one is 
\begin{equation}
{H^{RC}}_{g}= \int{\vec{B}.{\vec{J}}_{0}d^{3}x}
\label{19}
\end{equation}
From this definition of helicity , where ${\vec{J}}$ was simply replaced by ${\vec{J}}_{0}$, one is able to obtain the new expression for the variation of the helicity by
\begin{equation}
{{\Delta}H^{RC}}_{g}= \int{{\nabla}{\epsilon}.{\vec{J}}_{0}d^{3}x}=\int{{\nabla}.({\epsilon}{\vec{J}}_{0})d^{3}x}=\int{{\epsilon}{\vec{J}}_{0}.\vec{n}dS}
\label{20}
\end{equation}
Note that from expressions (\ref{20}) and (\ref{19}) it is possible to obtain 
\begin{equation}
{\vec{n}}.{\vec{J}}_{0}= {\vec{n}}.\vec{J}-{\vec{n}}.[{\nabla}{\times}\vec{B}]=0
\label{21}
\end{equation}
This expression and (\ref{20}) together lead to the conservation of the helicity in Riemann-Cartan spacetime given by  
\begin{equation}
{{\Delta}H^{RC}}_{g}= 0
\label{22}
\end{equation}
One must notice that the condition that leads to this conservation does not imply that the torsion vector $\vec{J}$ is now orthogonal to the static torsion loop plane, which solves the contradiction this would imply in teleparallel spacetime. In the spirit of this sectiona non-Riemannian loop magnetic metric can be obtained from the Letelier metric (\ref{1}), the only difference, however, is that now the vector $\vec{B}$ in the metric coeficient is a true magnetic field. This choice is very convenient since by makind use of Cartan's calculus of differential forms Letelier relation (\ref{2}) is identical to the magnetic equation where the torsion vector field $\vec{J}$ is now equivalent to a electric current. Therefore in this case our system is equivalent to a circular current carrying loop generating magnetic fields. In the next section makes use of another magnetic metric where now the metric is Riemannian and the MHD equations is consider to constrain the Cartan's equations. The advantage of considering $\vec{B}$ as a real magnetic field is that relation (\ref{2}) then implies a dynamo generating magnetic fileld from Cartan torsion. This idea has been fully sustained by De Sabbata and Gasperini \cite{13}and more recently by Opher and Wichoski \cite{14}.
\section{Riemannian magnetic metrics in MHD} 
In this section we shall consider the diagonal magnetic metric given by the following line element in cylindrical coordinates adequate to treat the geometry of tubes in plasma physics 
\begin{equation}
ds^{2}=dt^{2}-{B_{r}}^{2}dr^{2}-{B_{{\theta}}}^{2}r^{2}d{\theta}^{2}-{B_{z}}^{2}dz^{2}
\label{23}
\end{equation} 
where the basis one-forms are given by
\begin{equation}
{\omega}^{0}= dt
\label{24}
\end{equation} 
\begin{equation}
{\omega}^{1}={B_{r}}dr
\label{25}
\end{equation} 
\begin{equation}
{\omega}^{2}={B_{{\theta}}}r d{\theta}
\label{26}
\end{equation} 
\begin{equation}
{\omega}^{3}= {B_{z}}dz
\label{27}
\end{equation} 
By making use of the first Cartan's structure equation
\begin{equation}
T^{\alpha}= d{\omega}^{\alpha} +{{\omega}^{\alpha}}_{\beta}{\wedge}{\omega}^{\beta}
\label{28}
\end{equation} 
where the symbol ${\wedge}$ means the exterior product and ${{\omega}^{\beta}}_{\gamma}$ represents the connection one-form and $T^{\beta}$ represents the Cartan torsion two-form. All components of the magnetic field depends only upon the radial coordinate $r$. Since the metric is Riemannian the torsion forms vanish and this can be used together with the MHD equations for helical fields to constrain the geometry and to find out the Riemann curvature of the magnetic metrics. To able to accomplish this task we consider the following MHD equations in the steady-state case leading to the phenomelogical Maxwell equation 
\begin{equation}
k\vec{v}{\times}{\vec{B}}= {\nabla}{\times}\vec{B}
\label{29}
\end{equation}
where k is a constant. By considering the helical flow \cite{10} 
\begin{equation}
\vec{B}=({B_{r}},{B_{{\theta}}},{B_{z}})
\label{30}
\end{equation} 
\begin{equation}
\vec{v}=(0,{\omega}r,v)
\label{31}
\end{equation}
where ${\omega}$ and v are constants. These vectors used in equation (\ref{29}) yields the following conditions 
\begin{equation}
{\omega}rB_{z}= vB_{\theta}
\label{32}
\end{equation} 
\begin{equation}
-{B_{z}}'= kvB_{r}
\label{33}
\end{equation} 
\begin{equation}
{rB_{\theta}}'= -k{\omega}r^{2}B_{r}
\label{34}
\end{equation}
Here the upper prime represents derivation with respect to the radial coordinate r. By making use of the solenoid condition ${\nabla}.{\vec{B}}=0$ one obtains the following solution of the helical MHD flow
\begin{equation}
B_{r}= \frac{c_{1}}{r}
\label{35}
\end{equation} 
\begin{equation}
{B_{\theta}}=\frac{c_{2}}{r}-\frac{1}{2}k{\omega}c_{1}r
\label{36}
\end{equation} 
\begin{equation}
B_{z}= c_{3}-kvc_{1}lnr
\label{37}
\end{equation} 
To apply this MHD solution to constrain our Riemannian magnetic geometry one needs before to substitute the basis one forms into the Cartan equation (\ref{28}) which yields
\begin{equation}
T^{t}= {{\omega}^{t}}_{r}{\wedge}{\omega}^{r}+{{\omega}^{t}}_{\theta}{\wedge}{\omega}^{\theta}+
{{\omega}^{t}}_{z}{\wedge}{\omega}^{z}
\label{38}
\end{equation} 
\begin{equation}
T^{\theta}= d{\omega}^{\theta}+{{\omega}^{\theta}}_{r}{\wedge}{\omega}^{r}+{{\omega}^{\theta}}_{z}{\wedge}{\omega}^{z}+
{{\omega}^{\theta}}_{t}{\wedge}{\omega}^{t}
\label{39}
\end{equation}
\begin{equation}
T^{z}= {{\omega}^{z}}_{r}{\wedge}{\omega}^{r}+{{\omega}^{z}}_{\theta}{\wedge}{\omega}^{\theta}+
{{\omega}^{z}}_{t}{\wedge}{\omega}^{t}
\label{40}
\end{equation}
\begin{equation}
T^{r}= {{\omega}^{r}}_{z}{\wedge}{\omega}^{z}+{{\omega}^{r}}_{\theta}{\wedge}{\omega}^{\theta}+
{{\omega}^{r}}_{t}{\wedge}{\omega}^{t}
\label{41}
\end{equation}
where the only nonvanishing exterior derivative of the basis one-form is 
\begin{equation}
d{{\omega}^{\theta}}=[r{B'}_{\theta}+B_{\theta}]dr{\wedge}d{\theta}
\label{42}
\end{equation}
Substitution of (\ref{42}) into (\ref{39}) yields the only nonvanishing component of the connection one-form
\begin{equation}
{{\omega}^{\theta}}_{r}=\frac{[r{B'}_{\theta}+B_{\theta}]d{\theta}}{B_{r}}
\label{43}
\end{equation} 
where $c_{i}$ with $(i=1,2,3)$ are the integration constants. Substitution of Maxwell equation (\ref{34}) into expression (\ref{43}) yields the following constraint of MHD equations to the Riemannian geometry of magnetic metric
\begin{equation}
{{\omega}^{\theta}}_{r}= -k{\omega}r^{2}d{\theta}
\label{44}
\end{equation}
which from the second Cartan's structure equation
\begin{equation}
{R^{\alpha}}_{\beta}= d{{\omega}^{\alpha}}_{\beta}+{{\omega}^{\alpha}}_{\gamma}{\wedge}{{\omega}^{\gamma}}_{\beta}
\label{45}
\end{equation}
yields
\begin{equation}
{R^{\theta}}_{r}= d{{\omega}^{\theta}}_{r}
\label{46}
\end{equation} 
which along with the definition of the curvature two-form 
\begin{equation}
{R^{\alpha}}_{\beta}={R^{\alpha}}_{{\beta}{\gamma}{\delta}}{\omega}^{\gamma}{\wedge}{\omega}^{\delta}
\label{47}
\end{equation}
yields the following component for the Riemann tensor of the magnetic manifold
\begin{equation}
{R^{\theta}}_{r{\theta}z}= -2k{\omega}r
\label{48}
\end{equation}
which is equivalent to the expression for the Riemann curvature of the magnetic manifold
\begin{equation}
{R^{\theta}}_{r{\theta}z}= -2kv(r)
\label{49}
\end{equation}
This is the Riemannian curvature of the helical MHD flow. A similar relation between the curvature of intratube and the velocity of the flow has been obtained previously by Pelz \cite{15} in the context of vortex filament models. A particular case of the above Riemannian MHD metric maybe consider as
\begin{equation}
ds^{2}=dt^{2}-dx^{2}-{B_{y}}^{2}dy^{2}-dz^{2}
\label{50}
\end{equation}
where we have consider now the Cartesian retangular coordinates $(x,y,z)$ for the spatial part of the magnetic metric. Here the only nonvanishing component of the magnetic field is given by $B_{y}(x,y)$. By again making use of Cartan's calculus of differential forms yields the following equations
\begin{equation}
{{\omega}^{3}}_{2}= [{\nabla}{\times}\vec{B}]_{z}d{x}
\label{51}
\end{equation}
which in turn yields the following Bianchi identity
\begin{equation}
d{R^{3}}_{2}= d{{\omega}^{3}}_{2}={\partial}_{y}{\partial}_{z}[{\nabla}{\times}\vec{B}]_{z}dy{\wedge}dx{\wedge}dz
\label{52}
\end{equation}
By considering equation (\ref{29}) one obtains the equation
\begin{equation}
d{R^{3}}_{2}= d{{\omega}^{3}}_{2}={\partial}_{y}{\partial}_{z}[k\vec{v}{\times}\vec{B}]_{z}dy{\wedge}dx{\wedge}dz=0
\label{53}
\end{equation}  
which is consistent with the  Bianchi identity $d{R^{\alpha}}_{\beta}=0$ where d is the exterior derivative. In all the above example we found out an interesting interplay between the equations of plasma physics and the geometrical equations of the Cartan's calculus of differential forms. 
\section{conclusions}
A natural extension of the magnetic topology of torsion loops to investigate knots in MHD can be undertaken by generalising the static Letelier's torsion loops to time-dependent torsion loops. This extension unfortunately has been proved very difficult \cite{16} even by modern computation techniques using the OrtoCartan program. Non-Riemannian geometry and topology of torsion  curves is discussed. A new definition of the magnetic-like gravitational helicity is proposed. We show that the extension of teleparallel spacetime to Riemann-Cartan spacetime allows us to possible applications of the mathematics discussed here in astrophysical models may be proposed in near future. Solitonic equations in non-Riemannian background may also be considered in near future.
\section*{Acknowledgement}
I am very much indebt to P.S.Letelier, for helpful discussions on the subject of this paper, and to UERJ for financial support. 

\end{document}